\documentclass{sig-alternate-05-2015}

\usepackage{cite}
\usepackage{amssymb}
\usepackage{array}
\usepackage{url}
\usepackage{wasysym}
\usepackage{multirow}
\usepackage{url}
\usepackage[linesnumbered,ruled]{algorithm2e} 
\usepackage{color}
\usepackage{tikz}
\usepackage{multicol}
\usepackage{tabularx}

\usepackage{caption}
\usepackage{subcaption}
	\newcommand*\circled[1]{\protect\tikz[baseline=(char.base)]{
    		        \protect\node[shape=circle,draw,inner sep=2pt] (char) {#1};}}

\newtheorem{theorem}{Theorem}
\newdef{definition}{Definition}

\makeatletter
\def\@copyrightspace{\relax}
\makeatother

\begin{document}

\title{An Efficient and Robust Social Network \\ De-anonymization Attack}

\numberofauthors{3}
\author{
\alignauthor
G{\'a}bor Gy{\"o}rgy Guly{\'a}s\titlenote{Corresponding author.}\\
       \affaddr{INRIA, France}\\
       \email{gabor.gulyas@inria.fr}
\alignauthor
Benedek Simon\\
       \affaddr{BME, Hungary}\\
       \email{fleezeum@gmail.com}
\alignauthor
S{\'a}ndor Imre\\
       \affaddr{BME, Hungary}\\
       \email{imre@hit.bme.hu}
}

\date{13 October 2016}

\maketitle
\begin{abstract}
Releasing connection data from social networking services can pose a significant threat to user privacy. In our work, we consider structural social network de-anonymization attacks, which are used when a malicious party uses connections in a public or other identified network to re-identify users in an anonymized social network release that he obtained previously.

In this paper we design and evaluate a novel social de-anonymization attack. In particular, we argue that the similarity function used to re-identify nodes is a key component of such attacks, and we design a novel measure tailored for social networks. We incorporate this measure in an attack called Bumblebee. We evaluate Bumblebee in depth, and show that it significantly outperforms the state-of-the-art, for example it has higher re-identification rates with high precision, robustness against noise, and also has better error control.
\end{abstract}

%

\printccsdesc

\keywords{social network; privacy; anonymity; de-anonymization}

\section{Introduction}
\label{sec:intro}
The rise of social networking services has paved the way for businesses monetizing social data; however, abuses and malintents emerged as quickly as legal opportunities. In parallel, the deep integration of social networking services into everyday life provides a never seen possibility for governments to extend their spying capabilities \cite{prism}. This asymmetry of powers between the users versus business parties and governments continuously urges the investigation of privacy issues beyond positive uses of social networks.

In this paper we consider the problem how the graph structure of a network alone can be abused to match users in different social networks, which often leads to the violation of user privacy. More specifically, we focus on large-scale re-identification attacks when two large networks are aligned together in order to de-anonymize one of them. As a result, each node in the anonymized network is either connected to a known identity in another network or not, meaning that the identity of the node is revealed (or it is de-anonymized).

There is a myriad of motivations for such attacks. For example a business party can be financially motivated in the de-anonymization of freshly bought anonymized datasets in order to update its previous records with the new data. Meanwhile, gluing several communication networks together can be an attractive target for government agencies (e.g., email and telephone communications). Beside several sucessful attacks on social networks \cite{nar09, noseed13, yg13, noseed14, ada14, kl14, comm14, grh14}, it has also been shown that location data can be de-anonymized with social networks as a background knowledge \cite{mob12, ada14}.

The algorithm proposed by Narayanan and Shmatikov in 2009 (to which we refer as Nar) was the first to achieve large-scale re-identification in social networks \cite{nar09}. Their work applied comparison of nodes based on previously discovered matching of neighboring nodes, meaning that instead of comparing each node in background knowledge to all in the anonymized graph (i.e., global comparison), their algorithm compared each node only to a smaller set of others that had been close enough to existing mappings (local comparison). This reduced computational complexity of the problem significantly, thus allowing large-scale re-identification in networks consisting of even more than a hundred thousand nodes. The authors in their main experiment re-identified $30.8\%$ of co-existing nodes between a Twitter (224k nodes) and a Flickr (3.3m nodes) crawl with a relatively low error rate of $12.1\%$.

\begin{figure}[h!]
	\centering
	\includegraphics[width=.45\textwidth]{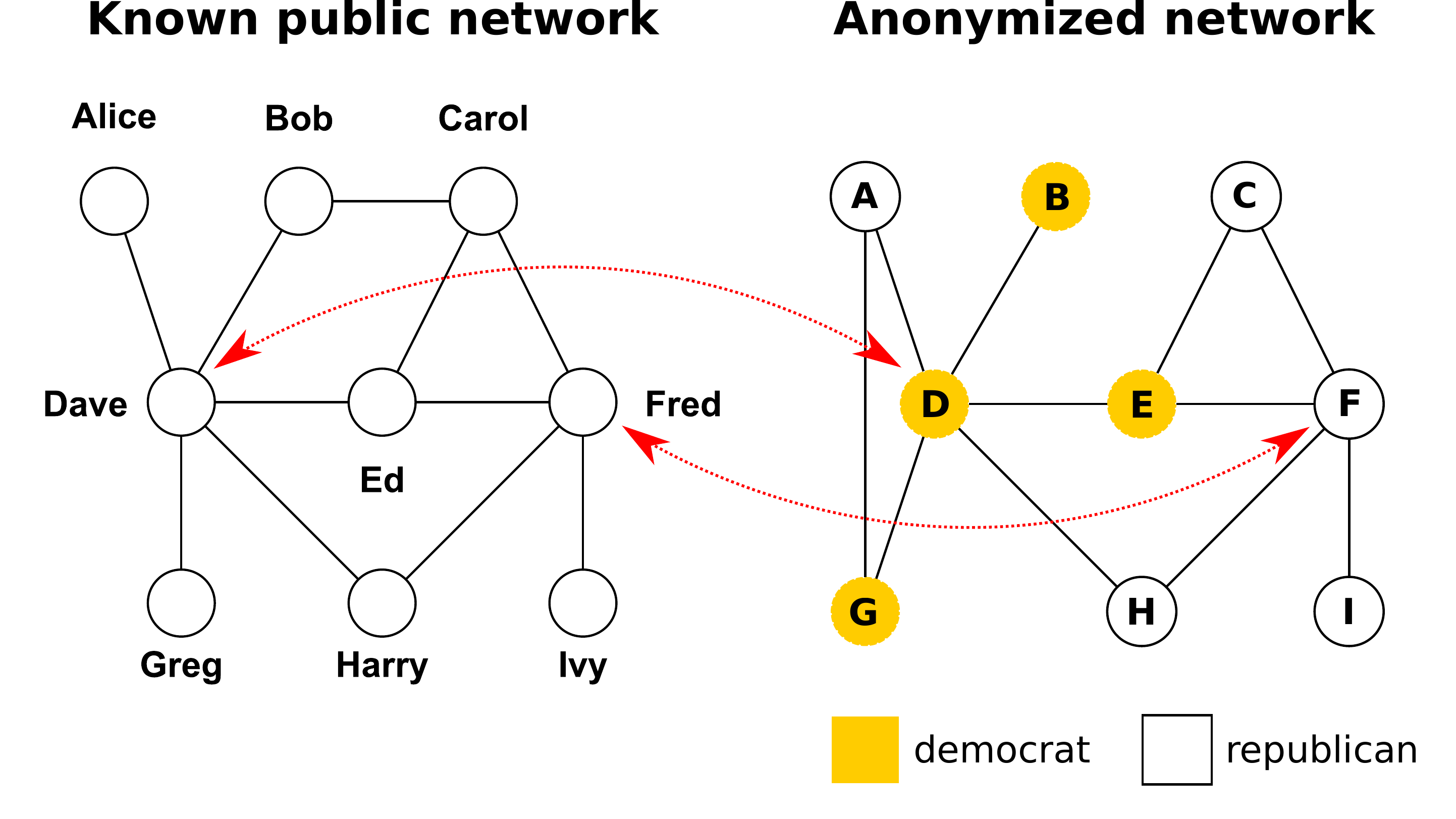}
    \caption{Datasets for the example of de-anonymization with an initial re-identification mapping between some nodes.}
    \label{fig:intro}
\end{figure}

Let us now illustrate how the Nar attack works on a simple example. An adversary obtains datasets as depicted in Fig. \ref{fig:intro}, a dataset of known identities that he uses as a background knowledge (left), and an anonymized/sanitized dataset (right). The attacker's goal is to to learn political preferences by structural de-anonymization. Initially, the attacker re-identifies (or maps) some nodes that stand out globally from the dataset (the seeding phase). In our example, two top degree nodes are re-identified: \texttt{Dave} $\leftrightarrow$ \texttt{D} and \texttt{Fred} $\leftrightarrow$ \texttt{F} -- as their degrees are outstandingly high and globally match each.

Then the attack continues by inspecting nodes related to the ones already re-identified (the propagation phase). The attacker picks a yet unidentified node in the background knowledge network. Let he first pick \texttt{Harry}, who has two neighbors already mapped: \texttt{Dave}, \texttt{Fred}. Nodes that could correspond to \texttt{Harry} in the anonymized dataset should also be neighbors of \texttt{D} and \texttt{F}, which are nodes \texttt{A}, \texttt{B}, \texttt{C}, \texttt{E}, \texttt{G}, \texttt{H}, and \texttt{I}. In the next step, similarity of these nodes are measured to \texttt{Harry}, which is calculated by dividing the number of their common mapped neighbors with their degree. This results a similarity score of $1$ for nodes \texttt{B}, \texttt{I}; $1/\sqrt{2}$ for nodes \texttt{A}, \texttt{C}, \texttt{G}; $2/\sqrt{3}$ for node \texttt{E}; and $2/\sqrt{2}$ for node \texttt{H}. As node \texttt{H} alone has the highest score, the attacker selects \texttt{Harry} $\to$ \texttt{H}. However, for precaution, the attacker uses the same technique to check which node would be the best match (in the public dataset) for \texttt{H}. If he gets \texttt{Harry} $\gets$ \texttt{H} in this case as well, the new mapping \texttt{Harry} $\leftrightarrow$ \texttt{H} is registered.

While the similarity metric used by Nar provides a good balance between correct and incorrect re-identification matches, it is biased towards low degree nodes. For instance, if the attacker picks \texttt{Ed} first, he cannot re-identify him correctly: node \texttt{H} would have a higher score than node \texttt{E} (with $2/\sqrt{3}$). Therefore we argue that selecting the right metric for comparison is a critical part of the algorithm, as it determines the performance and final limits of the attack.
We propose an alternative metric, for which we show by evaluation that it significantly improves the state-of-the-art. Our main contributions are the following:
\begin{itemize}
\item We create a novel structural re-identification algorithm called \emph{Bumblebee}, which is based on a new similarity metric specifically designed for the current re-identification task. We show that improving this critical point enhances several properties of the attack significantly. Parameters of Bumblebee allow a more fine-grained control than the Nar algorithm due to the larger score diversity.
\item We evaluate the performance, robustness of Bumblebee and its sensitivity of initialization.
We show that Bumblebee can reach significantly higher correct de-anonymization rates while having the same or lower error rate as of Nar. In our experiments, we observe an increase of $28\%$-$50\%$ of correct re-identifications while error remains the same. Inversely, error rates can also be decreased close to zero. We measured error rates below $0.2\%$ when having similar proportions of correct re-identification rates.
We also show that Bumblebee is significantly more robust to noise than Nar, as it achieves large-scale correct re-identification in cases where other attacks could not rise above $1\%$.
For seeding, we found that only $1-2$ seed nodes are enough for successful attacks, significantly smaller compared to previous works.
\item We compare the performance of Bumblebee to other modern de-anonymization attacks, and show that its performance is the most outstanding compared to the state-of-the-art, even against anonymized datasets as well.
\item We publish the source code of Bumblebee along our framework called Structural Anonymity Lab (\texttt{SALab}). Although we used earlier versions of it in our previous works, we now release \texttt{SALab} as an open source framework that supports the experimental analysis of re-identification attacks in social networks. It guarantees the reproducibility of our results.
\end{itemize}

The paper is organized as follows. In Section 2, we discuss the related work, and in Section 3, we provide the methodology of our evaluation. Section 4 provides the details of the Bumblebee algorithm, and in Section 5, we evaluate our algorithm and compare it to other ones as well. Finally, in Section 6, we conclude our work.

\section{Related Work}
\label{sec:related}
Several attacks have followed Narayanan and Shmatikov \cite{nar09}, and now we discuss the most relevant ones \cite{nar11, mob12, noseed13, yg13, noseed14, ada14, kl14, comm14, grh14}. These are heuristic approaches for solving the de-anonymization problem as finding the optimal solution is not computationally feasible \cite{noseed14}. The attacks may also differ in many details, albeit they share some common properties. Generally, they consist of two sequentially executed phases: one for initialization (seed phase with global node comparison), and another that extends re-identification in an iterative manner (propagation phase with local node comparison).

This is not a strict rule, some attacks run their propagation phase for seed generation \cite{noseed13, noseed14}, we call these self-starting attacks. Other attacks need seeds to start, which we call seed-based attacks \cite{nar09, nar11, mob12, yg13, ada14, kl14, comm14, grh14}.
Another important property of attacks is how they compare nodes. Some attacks use simple node properties for measuring node similarity \cite{nar09, nar11, yg13, kl14, grh14}, such as Nar uses cosine similarity, while others require deeper structural properties \cite{mob12, noseed13, ada14, noseed14}, like distance vector comparison where shortest path distances are considered from already mapped nodes \cite{mob12}.

Narayanan et al. published also the first attack following \cite{nar09}, that consisted of a simplified version of the original attack \cite{nar11}.
Srivatsa and Hicks showed on small datasets that location traces can be re-identified by matching them to social networks \cite{mob12}, and other algorithms were proposed to this problem by Ji et al. \cite{ada14}, verified on the same datasets. Pham et al. proposed an algorithm that converts spatiotemporal data into a social network on large datasets \cite{spt13}. Building upon their work, Ji et al. confirmed that location data can be efficiently re-identified with social networks \cite{mob14, noseed14}. Despite their original use, algorithms of \cite{mob12, ada14, noseed14} can also be used on attacking regular social networks.

The algorithm designed by Nilizadeh et al. uses other attacks as a base algorithm, and it exploits the cluster-oriented structure of the networks: runs the base re-identification algorithm first on the cluster structure, then inside them \cite{comm14}. In their evaluation they used the Nar algorithm. 
Pedarsani et al. proposed the first self-starting attack that required no seeding \cite{noseed13}. The works of Yartseva and Grossglauser \cite{yg13}, and the paper of Korula and Lattenzi \cite{kl14} contain simplified de-anonymization attacks in order to enable formal analysis of the algorithms. In our previous work we proposed Grasshopper \cite{grh14} (also referred to as Grh), a robust attack algorithm that works with very small error rates (typically less then $1\%$).

The first comparative evaluation of several structural de-anonymization attacks were first provided in \cite{secgraph16}. Seven algorithms in \cite{nar09, noseed13, mob12, noseed14, ada14, kl14, yg13} were selected based on generality, scalability and practicality, were compared regarding robustness of background knowledge and against anonymization. Ji et al found that some algorithms were the most prominent only conditionally; they concluded that none of the evaluated algorithms could be considered as generally better than others. To the best of our knowledge, we neither know about previously published attacks that have been proven to be better in general than the Nar algorithm. Thus, we considered the Nar algorithm as the baseline attack in our work, and we included Grasshopper also, as this attack has appealing properties, like fairly high re-identification rates in general with error rates almost at zero \cite{grh14}. However, we also provide comparison based on our own measurements against results published in \cite{secgraph16}.

To the extent of our knowledge, the only structural re-identification framework that is similar to \texttt{SALab}, is SecGraph \cite{secgraph16}. The benefit of SecGraph is that it implements several re-identification, anonymization and utility algorithms. Unfortunately, it does not implement several important functions that are necessary for large-scale experimentation, such as synthetic dataset perturbation (e.g., as in \cite{nar09}), seed generation algorithms, and detailed evaluation of results (e.g., saving every minute detail while attacks are executed). \texttt{SALab} had these functions and attacks implemented already, as previous versions of it has been used in multiple works: for the analysis of the importance of seeding \cite{seed14}, for measuring anonymity \cite{lta12, idsep15}, for evaluating the Grasshopper attack \cite{grh14}, and for analyzing numerous settings for adopting identity separation for protecting privacy \cite{idsep13, idsep15, thesis15}. Therefore we decided to choose \texttt{SALab} as our main tool.

\section{Notation and Methodology}
\label{sec:method}
We use the following notations in our work. Given an anonymized graph $G_{tar} = (V_{tar}, E_{tar})$ (target graph) to be de-anonymized by using an auxiliary data source $G_{src} = (V_{src}, E_{src})$ (where node identities are known), let $\tilde{V}_{src} \subseteq V_{src}, \tilde{V}_{tar} \subseteq V_{tar}$ denote the set of nodes mutually existing in both. Ground truth is represented by the mapping $\mu_{G} : \tilde{V}_{src} \rightarrow \tilde{V}_{tar}$ denoting the relationship between coexisting nodes. Running a deterministic re-identification attack on ($G_{src}$, $G_{tar}$) initialized by a seed set $\mu_0 : {V}_{src}^{\mu_0} \rightarrow {V}_{tar}^{\mu_0}$ results in a re-identification mapping denoted as $\mu : {V}_{src}^{\mu} \rightarrow {V}_{tar}^{\mu}$. Nodes are denoted as $v_i$, its set of immediate neighbors are denoted as $V_i \subseteq V$, and its neighbors of neighbors are denoted as $V_i^2 \subseteq V$.

In our work, we use simple measures for assessing the extent of what the attacker could learn from $\mu$. The \emph{recall rate} reflects the ratio of correct re-identifications: it is the number of correct re-identifications divided by the number of mutually existing nodes ($\tilde{V}$). The \emph{error rate} is the proportion of incorrectly re-identified nodes compared to $\tilde{V}$; however, as there might be erroneous mapping outside $\tilde{V}$, this leaves the possibility of the error rate to emerge above $100\%$. We also use \emph{precision}, which is the proportion of recall divided by the mapping size ($|\mu|$).

Data preparation and simulation parameters are key parts of methodology. We used multiple, large datasets with different characteristics in order to avoid related biases. We obtained two datasets from the SNAP collection \cite{snap}, namely the Slashdot network (82k nodes, 504k edges) and the Epinions network (75k nodes, 405k edges). The third dataset is a subgraph exported from the LiveJournal network crawled in 2010 by us (66k nodes, 619k edges). Datasets were obtained from real networks which assures that our measurements were being realistic.

We used the perturbation strategy proposed by Narayanan and Shmatikov \cite{nar09} to generate synthetic graph pairs for our experiments. Their method only uses deletion, thus the original graph acts as a ground truth between $G_{src}, G_{tar}$. Their algorithm makes two copies of the initial graph as $G_{src}, G_{tar}$, then performs independent deletions of nodes to achieve the desired fraction of overlap (denoted as $\alpha_{v}$), and then deletes edges independently to set the size of overlapping edges (to $\alpha_{e}$). Overlap parameters $\alpha_{v}, \alpha_{e}$ are measured by the Jaccard similarity\footnote{$\text{JaccSim}(U, V) = \frac{|U \cap V|}{|U \cup V|}.$}. These parameters control the level of noise in the data, therefore they are also a representation of the strength of anonymization and the quality of the adversarial background knowledge. Unless noted otherwise, we used the setting of $\alpha_{v}=0.5$, $\alpha_{e}=0.75$, where a significant level of uncertainty is present in the data and thus harder to attack.

The default simulation settings were set as follows. In each experiment we created two random perturbations, and run simulations once, as algorithms provided deterministic results when we used the same seed set (e.g., see \texttt{top} below). When seeding was random, we run the attacks multiple times. Nar and Grh both have an important parameter $\Theta$ that controls the ratio of correct and incorrect matches (greediness): the lower $\Theta$ is the less accurate mappings the algorithm will accept. As we measured fairly low error rates even for small values of $\Theta$, we have chosen to work with $\Theta=0.01$ for both algorithms.

The seeding method and size have been showed to have a strong effect on the overall outcome of de-anonymization \cite{seed14}. In \cite{seed14} we provided details for various methods, and highlighted seeding methods that were top performers on large networks, regardless of the network structure, e.g., nodes with the highest degree and betweenness centrality scores. Considering these results, we applied two seeding methods. Our prime choice was selecting an initial mapping between top degree nodes (denoted as \texttt{top}), as this method proved to be the most effective in previous measurements in terms of run-time and smallest seed set size \cite{seed14}. It can also be used in real attacks, as a weighted graph matching technique has been proposed for \texttt{top} seeding in \cite{nar11}. Seed sizes were typically $50-200$ for \texttt{top}. In some cases, we also considered random seed selection with high degree nodes, where nodes are selected from the top $1\%$ by degree (denoted as \texttt{random.01}).

\pagebreak

\section{The Bumblebee Attack}
\label{sec:bumblebee}

\subsection{The Motivation}
Biases of the similarity measure of Nar and Grh, such as we discussed in the introduction, motivated the invention of the Bumblebee attack (Blb). We now illustrate these biases and the new scoring function we propose on the source node $v_i \in V_{src}$ and possible mapping candidates denoted as $v_j \in V_{tar}$. The similarity measure in Nar is a simplification of the cosine similarity \cite{nar09} as

\begin{equation}
\label{eq:narsim}
\text{NarSim}(v_i, v_j) = \frac{|V_i \cap V_j|}{\sqrt{|V_j|}},
\end{equation}

where only a division with $\sqrt{|V_i|}$ is omitted, as this would be constant while comparing $v_i$ to any other nodes (n.b. $|V_i \cap V_j|$ is the number of common mapped neighbors). Unfortunately \eqref{eq:narsim} shows biases towards low degree nodes when the degree of $v_i, v_j$ differ significantly.

\begin{figure*}
\centering
\begin{minipage}[c][][t]{.6\textwidth}
	\vspace*{\fill}
  	\centering
	\includegraphics[width=\textwidth]{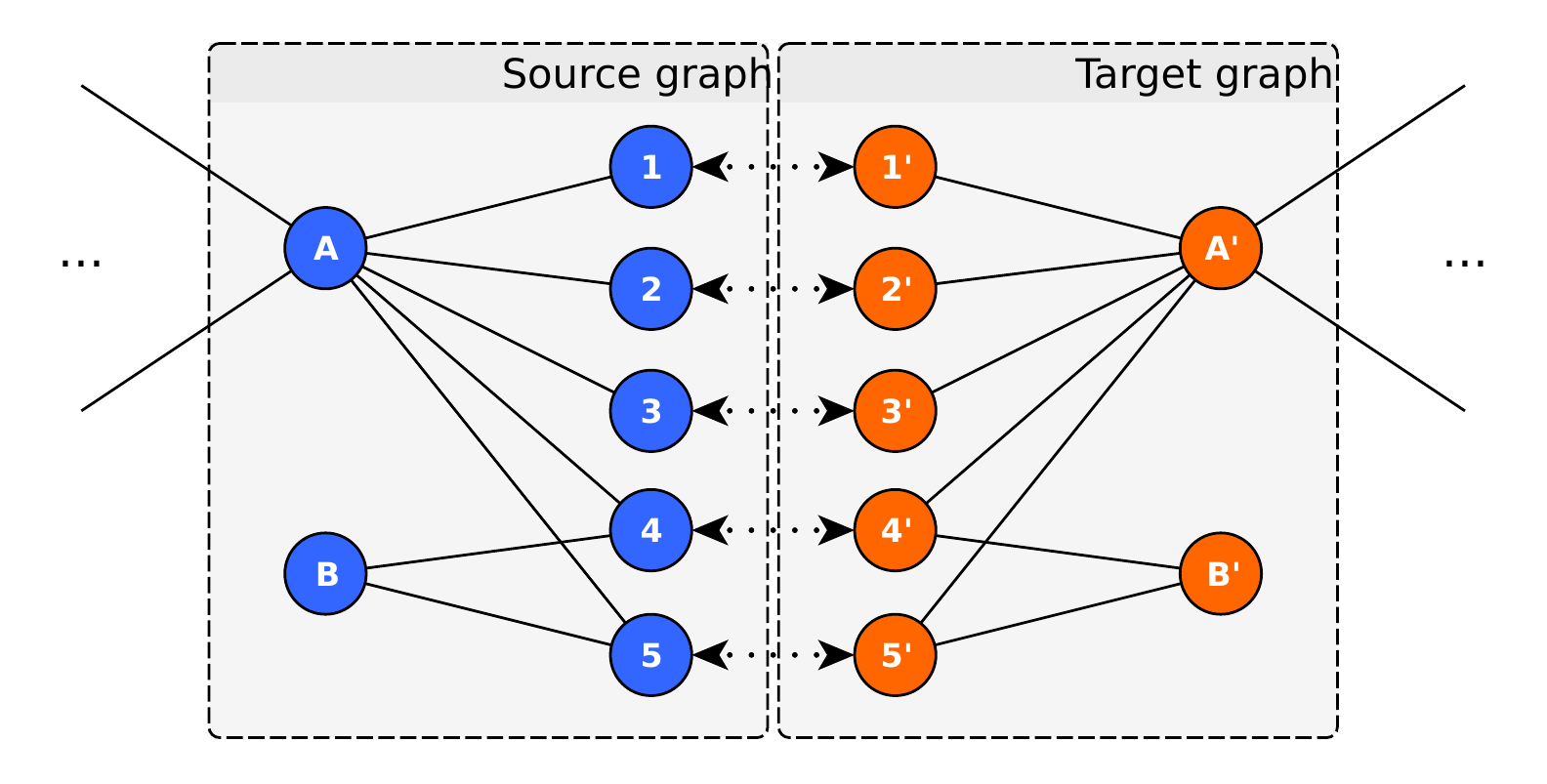}
	\subcaption{Here, $\text{deg}(v_A) = \text{deg}(v_{A'}) = 100$, and $\text{deg}(v_B) = \text{deg}(v_{B'}) = 2$ and $5$ mappings already exist in $\mu$.}
	\label{fig:motivation:example}
\end{minipage}
\:
\begin{minipage}[c][][t]{.3\textwidth}
  \vspace*{\fill}
  \centering
	\begin{tabular}{c|c|c||c}
		NarSim & $A'$ & $B'$  & ?\\ \hline
		$A$ & $\frac{5}{\sqrt{100}}$ & $\frac{2}{\sqrt{2}}$ & $B'$ \\ \hline
		$B$ & $\frac{2}{\sqrt{100}}$ & $\frac{2}{\sqrt{2}}$ & $B'$ \\
	\end{tabular}
    \subcaption{NarSim decisions}
    \label{fig:motivation:narsim}
    \quad
    	\centering
	\begin{tabular}{c|c|c||c}
		GrhSim & $A'$ & $B'$  & ?\\ \hline
		$A$ & $5$ & $2$ & $A'$ \\ \hline
		$B$ & $2$ & $2$ & $\text{N/A}$ \\
	\end{tabular}
    \subcaption{GrhSim decisions with $\forall \omega_i = 1$ ($\omega_i \in \omega $)}
    \label{fig:motivation:grhsim}
    \:
    	\centering
	\begin{tabular}{c|c|c||c}
		BlbSim & $A'$ & $B'$  & ?\\ \hline
		$A$ & $5$ & $0.89$ & $A'$ \\ \hline
		$B$ & $0.89$ & $2$ & $B'$ \\
	\end{tabular}
    \subcaption{BlbSim  decisions with $\delta=0.5$}
    \label{fig:motivation:blbsim}
\end{minipage}
\caption{Example for demonstrating biases in the similarity measures.}
\label{fig:motivation}
\end{figure*}

We demonstrate this on the example as shown in Fig. \ref{fig:motivation:example}. Here, we assume a situation where each algorithm has to make a decision, having two nodes, namely $v_A,v_B$, to find the correct mapping for. We assume a significant difference in node degrees, having $\text{deg}(v_A) = \text{deg}(v_{A'}) = 100$, and $\text{deg}(v_B) = \text{deg}(v_{B'}) = 2$, and also have $5$ mappings already existing in $\mu$ of which $2$ overlap. Given this situation, we calculated similarities and the resulting decisions, see it in Fig. \ref{fig:motivation:narsim}. As NarSim penalizes high degree nodes, the re-identification algorithm cannot re-identify $v_A$.

The Grasshopper algorithm uses a similarity measure that weights mappings \cite{grh14}:

\begin{equation}
\label{eq:grhsim}
\begin{split}
\text{GrhSim}(v_i, v_j)
&= \sum_{\substack{\forall (v', v'') \in \{V_i \cap V_j\} \\ \mathrm{where~} \mu(v') = v''}} \omega[v']\\
&= \sum_{\substack{\forall (v', v'') \in \{V_i \cap V_j\} \\ \mathrm{where~} \mu(v') = v''}} \Bigg( \frac{|V' \cap V''|}{\sqrt{|V'| \cdot |V''|}} + 1 \Bigg)
\end{split}
\end{equation}

where $\omega$ is the weight of each mapping in $\mu$. Both $\omega$, $\mu$ are updated at the end of every iteration. This measure is biased towards high degree nodes as the sum increases faster with a higher number of connections rather than having a smaller number of high degree adjacent nodes. Thus, GrhSim cannot discover node $v_B$ as we demonstrate this with the similarity matrix shown in Fig. \ref{fig:motivation:grhsim} (all weights are equal: $\forall \omega_i = 1$).

This leads to the need of a measure which yields higher values when degree values are similar. We propose the following similarity measure to remedy the situation:

\begin{equation}
\label{eq:blbsim}
\text{BlbSim}(v_i, v_j) = \bigg|V_i \cap V_j\bigg| \cdot \bigg(\min \bigg(\frac{|V_i|}{|V_j|},\frac{|V_j|}{|V_i|}\bigg)\bigg)^{\delta},
\end{equation}

where $\delta$ is a parameter that can be chosen freely as $\delta \in [0, 1]$. This similarity measure is more balanced than the previous ones when node degrees differ, as it is symmetric and yields higher score when node degrees are closer to each other. BlbSim can correctly re-identify both $v_A$ and $v_B$ as in Fig. \ref{fig:motivation:blbsim}.

\vfill\eject

\subsection{Evaluation of BlbSim}
Let us now evaluate if BlbSim is in fact better than NarSim, and if so, under what conditions. When two nodes are compared with the discussed similarity metrics, we use the already discovered mappings in $\mu$ to glue the source and target networks together. Here, by using the mappings as additional edges, $v_i \in V_{src}$ and all possible candidates $v_j \in V_{tar}$ are reckoned as neighbors of neighbors in the same graph. In this setting, we assume that if nodes $v_i$ and $v_j$ correspond to each other ($\exists \mu_G(v_i)=v_j$) then $v_i$, $v_j$ should be structurally similar (to some extent). An appropriate similarity measure $S(\cdot)$ should yield the highest score for $S(v_i, v_j)$ than for all other possible candidates. We can also describe this with the following correct re-identification decision criteria:

\begin{equation}
\label{eq:reidcrit}
\text{S}(v_i, v_j) > \text{S}(v_i, v'_j) + \Theta,
\end{equation}

where $v_i \in V_{src}, v_j, v'_j \in V_{tar}, v_j \neq v'_j$. The ratio between correct and false decisions depend on the value chosen for $\Theta \in [0, \infty)$.

Due to the way existing mappings are used, this is similar to the problem when we compare two nodes in the same graph $v_A, v_B \in V$. As we expected correct decisions in case of re-identification (eq. \eqref{eq:reidcrit}), we expect the similarity measure $S(\cdot)$ to give higher score for $v_A$ to itself, then of any other node $v_B \in V \setminus \{v_A\}$, which is written as

\begin{equation}
\label{eq:simexp}
\text{S}(v_A, v_A) > \text{S}(v_A, v_B) + \Theta
\end{equation}

Let us call this the \emph{graph node structural comparison problem}.

\begin{theorem}
\label{th:1}
Given the graph node structural comparison problem, having $\Theta=0$ and $\delta \geq 1/2$, $\text{BlbSim}(\cdot)$ leads to a higher number of correct decisions than $\text{NarSim}(\cdot)$.
\end{theorem}

We provide the proof for this theorem in the Appendix. While we restricted ourselves to some parameter settings in Theorem \ref{th:1}, later we show with experiments that $\Theta$ and $\delta$ parameters can be used together to achieve settings where the attack will exhibit interesting properties, e.g., providing high recalls or having extremely low error rates. Furthermore, as the scoring schemes output a different distribution, for BlbSim parameter $\Theta$ gives a significantly richer control of greediness (cf. in Fig. \ref{fig:robustness:thetas}).

\subsection{Attack Description}
\begin{algorithm}
	\scriptsize
    \caption{\textsc{Propagate}}
    \label{algo:propagate}
	\KwData{$G_{src}, G_{tar}, \mu$}
	\KwResult{$\mu, \Delta$}
	$\Delta \gets 0$\;
    \For{$v_{src} \in V_{src}$}{
		$S \gets $ \textsc{Score(}$G_{src}, G_{tar}, v_{src}, \mu$\textsc{)}\;\label{line:firstscore}
		\If{\textsc{Ecc}$(S.\textsc{values}()) < \Theta$}{\label{line:eccentricity}
			\textsc{continue}\;
		}
		$v_c \gets $ \textsc{Random}$($\textsc{Max}$(S))$\;
		$S_r \gets $ \textsc{Score(}$G_{tar}, G_{src}, v_c, \mu^{-1}$\textsc{)}\;\label{line:reverse1}
		\If{\textsc{Ecc}$(S_r.\textsc{values}()) < \Theta$}{
			\textsc{continue}\;
		}
		$v_{rc} \gets$ \textsc{Random}$($\textsc{Max}$(S_r))$\;\label{line:reverse2}
		\If{$v_{src} = v_{rc}$}{
			$\mu[v_{src}] \gets v_c$\;\label{line:newmapping}
			$\Delta \gets \Delta + 1$\;\label{line:convergence}
		}
    }
\end{algorithm}

\begin{algorithm}
    \scriptsize
    \caption{\textsc{Score}}
    \label{algo:score}
	\KwData{$G_{src}, G_{tar}, v_{src}, \mu$}
	\KwResult{$S$}
	$S \gets$ \textsc{Dict}$(\forall v \in V_{tar}: v \to 0)$\;
	\For{$v_i \in G_{src}.\textsc{nbrs}(v_{src})$}{\label{line:non1}
		\If{$\nexists \mu(v_i)$}{
			\textsc{continue}\;
		}
		\For{$v_j \in G_{tar}.\textsc{nbrs}(\mu(v_i))$}{
			\If{$\exists \mu^{-1}(v_j)$}{\label{line:non2}
				\textsc{continue}\;
			}
			$r_1 = \textsc{deg}(v_{src})/\textsc{deg}(v_j)$\;
			$r_2 = \textsc{deg}(v_j)/\textsc{deg}(v_{src})$\;
			$S[v_j] \gets S[v_j] + (\textsc{min}(r_1, r_2))^\delta$\;
		}
	}
\end{algorithm}

We now discuss how the complete attack works based on BlbSim, for which we provide the pseudocode as in Algo. \ref{algo:propagate} and \ref{algo:score}. After obtaining the seed set ($\mu_0$), the attack starts by iteratively calling \textsc{Propagate}, and it stops when no new mappings can be registered (when $\Delta = 0$). 
In each propagation step all nodes in $V_{src}$ are checked for new mappings. Existing mappings are visited again, as this allows correction after discovering more mappings. Then we calculate the BlbSim between $v_{src}$ and all possible candidates in \textsc{Score} by using $\delta$ (Algo. \ref{algo:propagate} line \ref{line:firstscore}). \textsc{Score} selects candidates by iterating unmapped neighbors of neighbors through $\mu$ (Algo. \ref{algo:score} lines \ref{line:non1}-\ref{line:non2}).

If there is an outstanding candidate $v_c$ (Algo. \ref{algo:propagate} line \ref{line:eccentricity}), we do a reverse checking to decrease false positives. We test if we swap the graphs and invert the mapping, would $v_{src}$ be the best candidate for $v_c$ (Algo. \ref{algo:propagate} lines \ref{line:reverse1}-\ref{line:reverse2}). If it is, we register the new mapping and indicate convergence (Algo. \ref{algo:propagate} lines \ref{line:newmapping}, \ref{line:convergence}). We say that a node with maximum score in $S$ is outstanding if its similarity score is higher than all others at least by $\Theta$:

\begin{equation}
\label{eq:ecc}
\textsc{Ecc}(S) = \frac{\max(S) - \max(S \setminus \{\max(S)\})}{\sigma(S)} \geq \Theta.
\end{equation}

An off-the-shelf working version of this attack is available in \texttt{SALab}. Bumblebee inherited some parts of Nar which we found to be natural and good design choices, such as candidate selection through mappings and reverse checking to eliminate false positives. However, we use a greedy convergence criteria triggered by even the smallest changes (Algo. \ref{algo:propagate} line \ref{line:convergence}), and we redesigned the scoring function (\textsc{Score}) for eliminating biases. Albeit we use the same eccentricity function as Nar, due to the wider scoring diversity of BlbSim, this plays a more important role in Blb, and allows a more fine-grained control of the algorithm (cf. in Fig. \ref{fig:robustness:thetas}).

\section{Evaluation}
\label{sec:evaluation}
In this section we evaluate the Bumblebee algorithm, and compare it to other attacks. Although we use multiple settings of Bumlbebee, parameters $\Theta$ and $\delta$ do remain free to choose. Best settings may depend on several factors, such as the size and structure of the dataset under analysis. We use the notation as Blb$(\delta)$ (when $\Theta$ is fixed) or as Blb$(\Theta, \delta)$.

As we discussed before, we use \texttt{SALab} which enables analyzing various aspects of large-scale structural attacks. We also make it available at \url{https://github.com/gaborgulyas/salab}; we hope it will contribute to further research helping to have a better understanding of structural anonymity.

Our simulations were run on a commodity computer with an Intel Xeon 3.6 GHz processor and 8 GB RAM.

\subsection{Experimental Analysis of $\delta$ and $\Theta$}
\label{sec:parameters}
\begin{figure}
	\centering
	\includegraphics[width=0.5\textwidth]{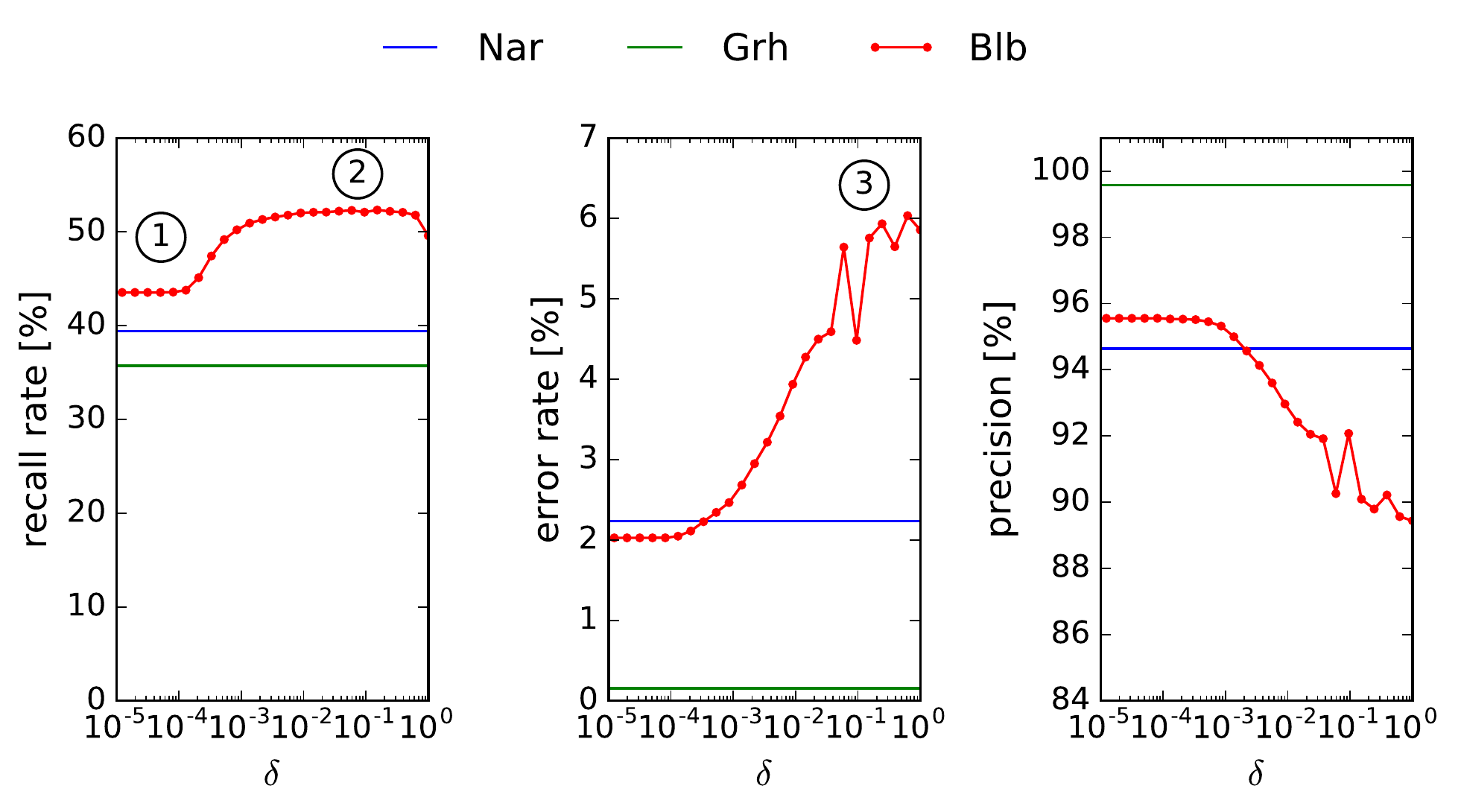}
	\caption{Effect of the $\delta$ parameter on results (Slashdot dataset).}
	\label{fig:delta2}
\end{figure}

First we evaluate how parameters $\delta$ and $\Theta$ influence the overall outcome. We used \texttt{top} seeding with $200$ nodes and $\Theta = 0.01$, meaning a greedy setting in case of Nar \cite{nar09} and Grh \cite{grh14} (we provide analysis of $\Theta$ for Bumblebee later), safe enough for all algorithms to provide high recall rates \cite{grh14, seed14}. Results for a wide range of $\delta$ values in the Slashdot dataset are visible in Fig. \ref{fig:delta2}. Due to space limitations, we select results of one network for demonstration; however, until noted otherwise, we were getting similar results in the other networks as well.

Interestingly, we observed transition phases of the recall and error rates as a function of $\delta$, which divided results into three interesting parameter subclasses for Blb. We used these classes for analyzing parameter $\Theta$. In the first class (denoted with \circled{1}) we observed lower recall rates and low error. We use such a setting as $\delta=0$ and denote it as Blb$(0)$. This conservative setting makes decisions only based on the number of common mappings without considering any other information. Another classes are where the recall rates plateau after increasing significantly (\circled{2}). 
Here we selected two settings, one when recall peaked but error is relatively low, and the other when the error is also peaked (\circled{3}). We observed that the recall rate usually at its highest at $\delta=0.005$ while error rate is still lower, thus we selected Blb$(.005)$, and finally we selected Blb$(.5)$ also.

\begin{figure}
	\centering
	\includegraphics[width=0.5\textwidth]{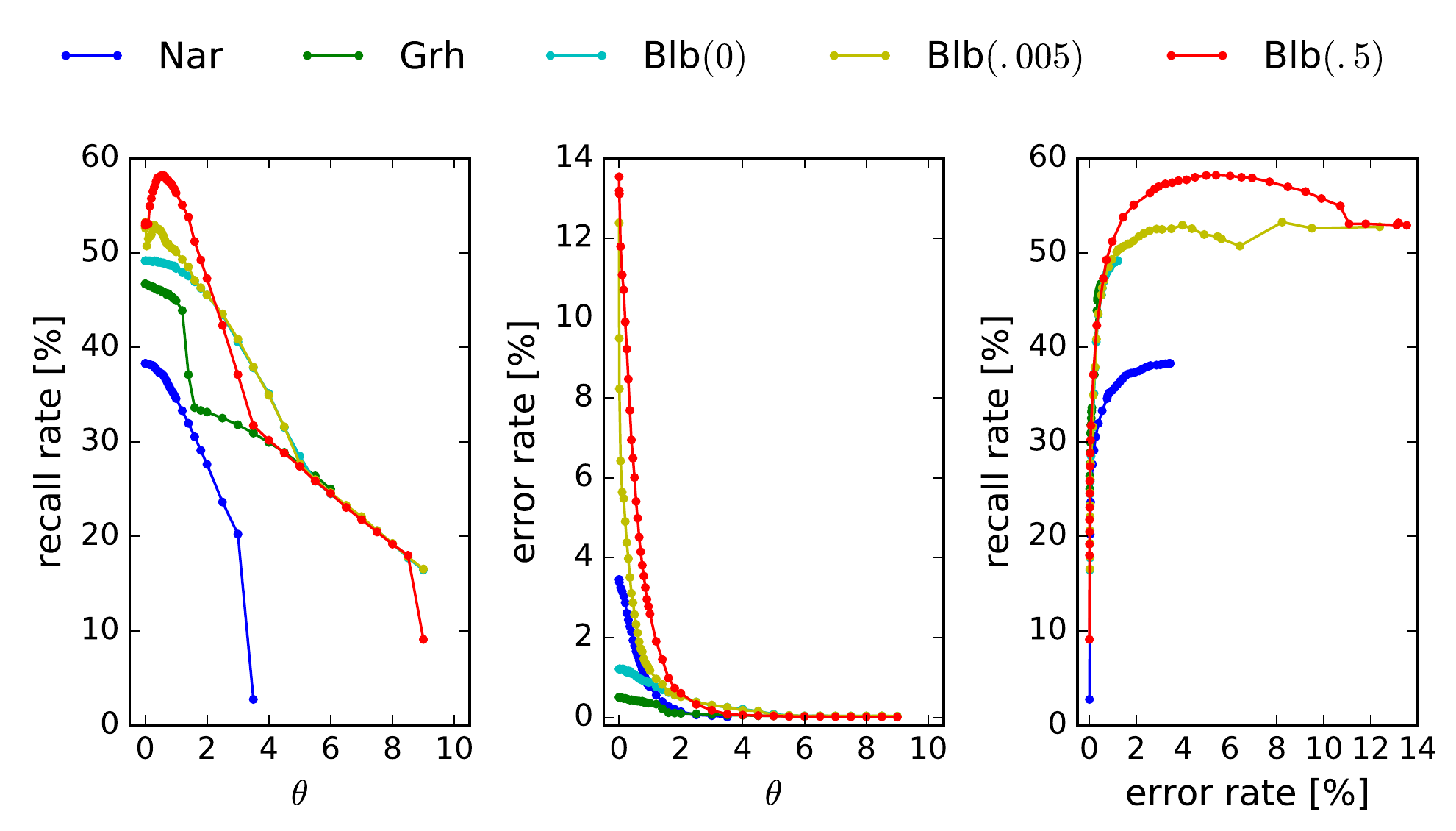}
	\caption{Effect of the $\Theta$ parameter for different algorithms in the LiveJournal dataset.}
	\label{fig:theta:lj}
\end{figure}

We provide comparison on the effect of the $\Theta$ parameter for the LiveJournal dataset in Fig. \ref{fig:theta:lj}. What clearly stands out from these results, that Blb could achieve the highest recall rates, while it also had higher errors for the same $\Theta$ settings compared to Nar and Grh. Furthermore, the conservative setting of Blb$(0)$ managed to achieve higher recall than Grh with similarly low error, and in most cases higher recall as Nar, but with a lower error rate. Another interesting property of Bumblebee is that it achieved high recall rates as Nar with extremely low error, e.g., Blb$(\Theta=4.0,\delta=0.5)$ recall = $29.91\%$, error = $0.15\%$; in other cases it achieved a significantly better trade-off between recall-error rates, as the error/recall results suggest. Comparing the recall-error results additionally shows that results provided by Grh, Blb$(0)$ are subclasses of the results of Blb$(.5)$. The only difference is they provide the same (recall, error) results at different values of $\Theta$. Finally, Blb$(.005)$ is clearly worse than Blb$(.5)$.

From these results, we conclude that both parameters of Bumblebee enable the control of greediness in the algorithm by taking values as $\Theta \in [0, \ldots, \infty)$, $\delta \in (0, ..., 1]$. Conservative (low error) settings are typically with high $\Theta$ and low $\delta$, greedy settings are with the opposite settings. In addition, parameters can also be used compensate each other, like as Blb$(1, .5)$ provided high recall with low error in Fig. \ref{fig:theta:lj}. We used these measurements to set parameters for further experiments.

\subsection{Analysis of Seeding Sensitivity}
\begin{figure*}[ht!]
	\centering
	\includegraphics[width=0.95\textwidth]{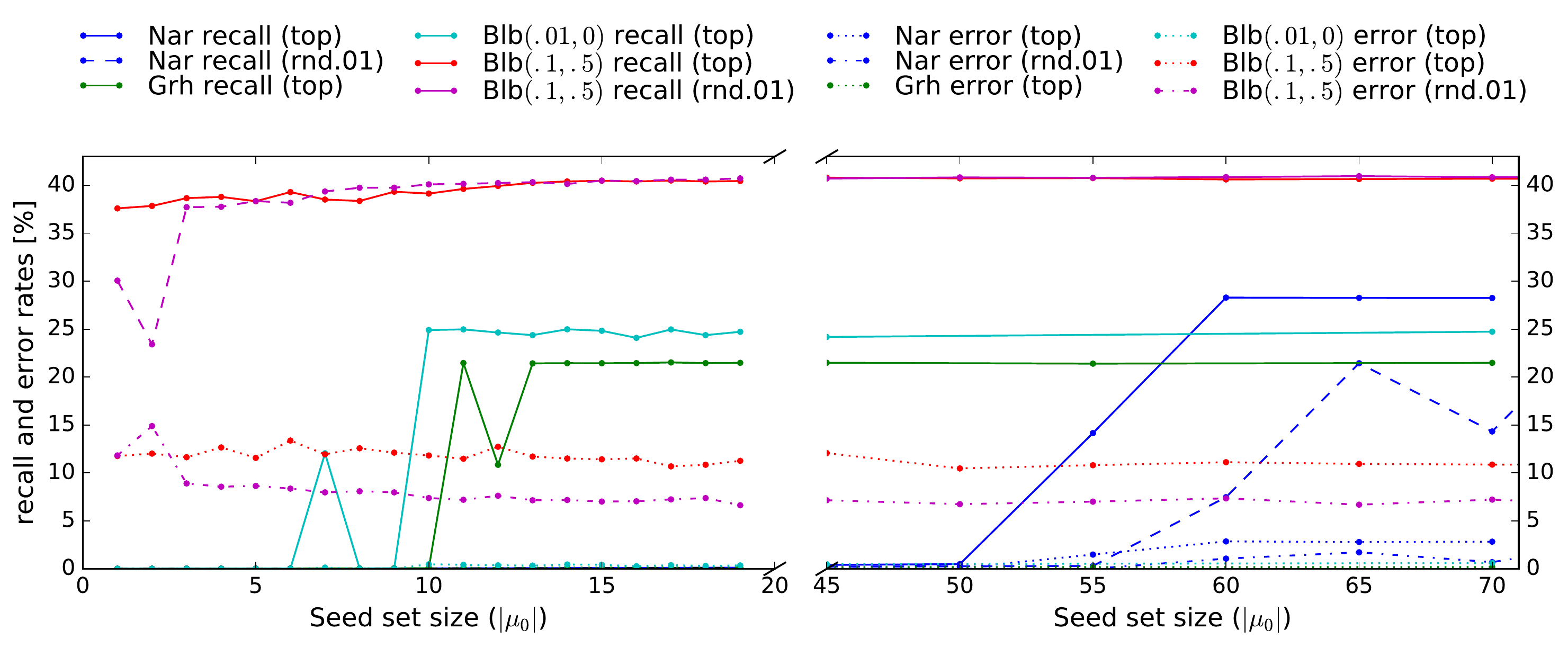}
	\caption{Seed sizes of \texttt{top} and \texttt{random.01} required for large-scale re-identification (Epinions dataset).}
	\label{fig:seeding:ep}
\end{figure*}

The number of required seeds is an important aspect of seed-based attacks, as these need to be provided before starting propagation, or in another sense, before starting the attack. Thus, the smaller the seed size is, the better. We consider the minimum required number of seeds for which the attack is stable: observed recall and error rates only vary minimally and the recall rate reaches its maximum \cite{seed14}. We provide an example on the minimum number of seed nodes with \texttt{top} in Fig. \ref{fig:seeding:ep} on Epinions dataset. Beside getting better recall and error rates than others, some Bumblebee variants managed to start with only a single seed node.


In other networks $1$-$2$ seed nodes were enough to achieve large-scale re-identification with a higher error rate, which dropped when the number of seeds reached $5$-$6$. We observed similar results when relaxing the seed criteria to \texttt{random.01}, multiple successful seeding attempts with a single random seed for Bumblebee. This means top $1\%$ of high degree nodes, random selection from approx. a set of $200$-$300$ nodes.

To the best of our knowledge, these results are superior to existing previous works, which reported minimum number of seeds at $16$ seed nodes \cite{comm14}, or $8-14$ seed nodes \cite{grh14}. In addition, relaxed seed criteria means easier attacks: one randomly selected node from the top $1\%$ as a seed is an easily doable task for an attacker; even with some retries or collecting a handful of seeds in order to achieve large-scale propagation.


\subsection{Robustness}
The quality of the background knowledge, or the overlap between $G_{src}$ and $G_{tar}$, has a strong influence on the limits of re-identification. For example, the Grh algorithm has typically higher recall rates than Nar when the overlap is smaller \cite{grh14}. In order to investigate such biases, we measured the performances of the algorithms with different perturbation settings. We note that almost all overlaps in these measurements are significantly lower than in experiments of Section \ref{sec:comparison}.

\begin{figure}
    \centering
    \begin{subfigure}[b]{0.32\textwidth}
		\centering
		\includegraphics[width=\textwidth]{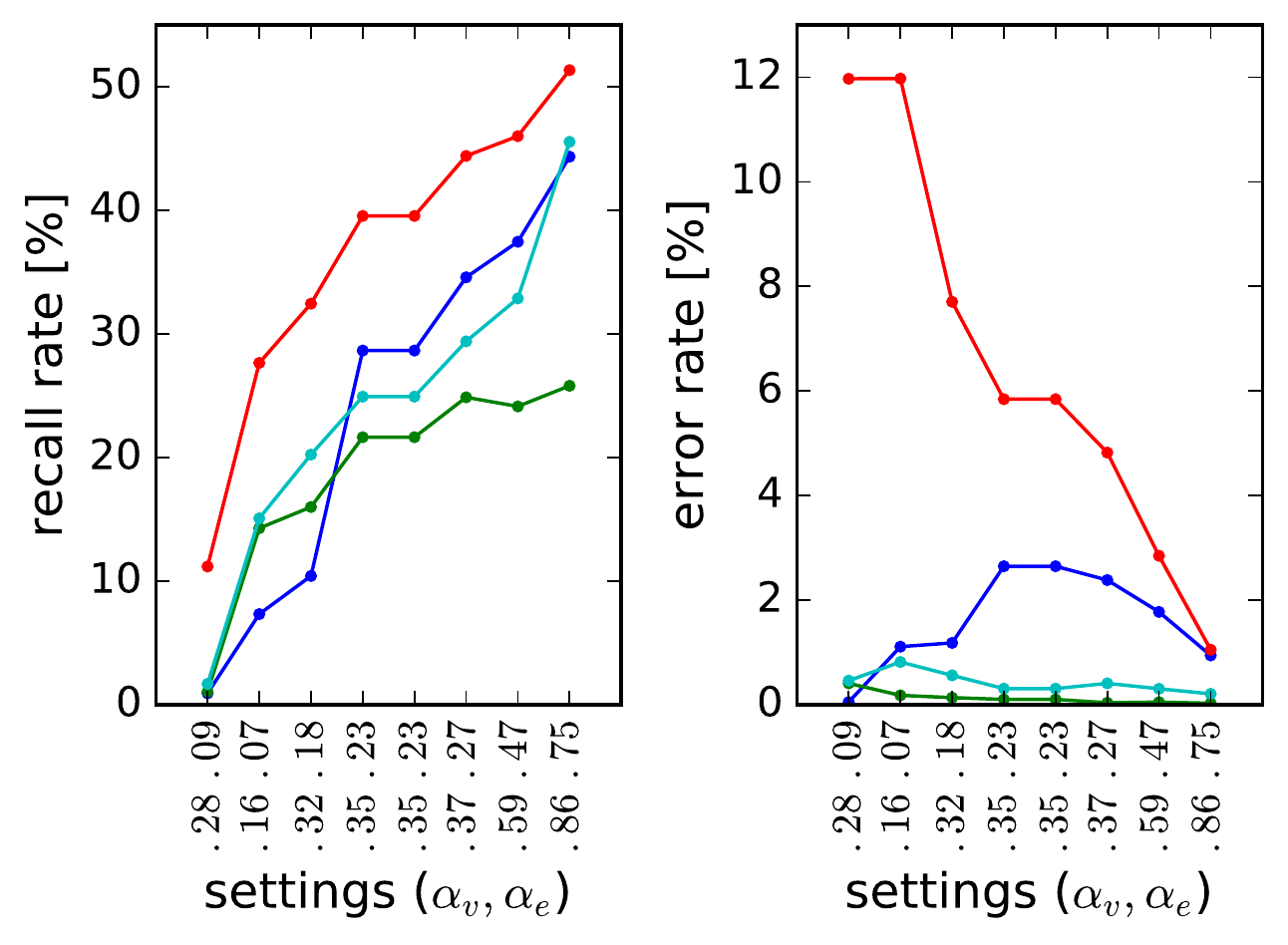}
		\caption{Performances for different perturbation settings. (overlap values are as actually measured, not as set for perturbation)}
		\label{fig:robustness:perts}
    \end{subfigure}
    ~
    \begin{subfigure}[b]{0.14\textwidth}
    		\centering
		\includegraphics[width=\textwidth]{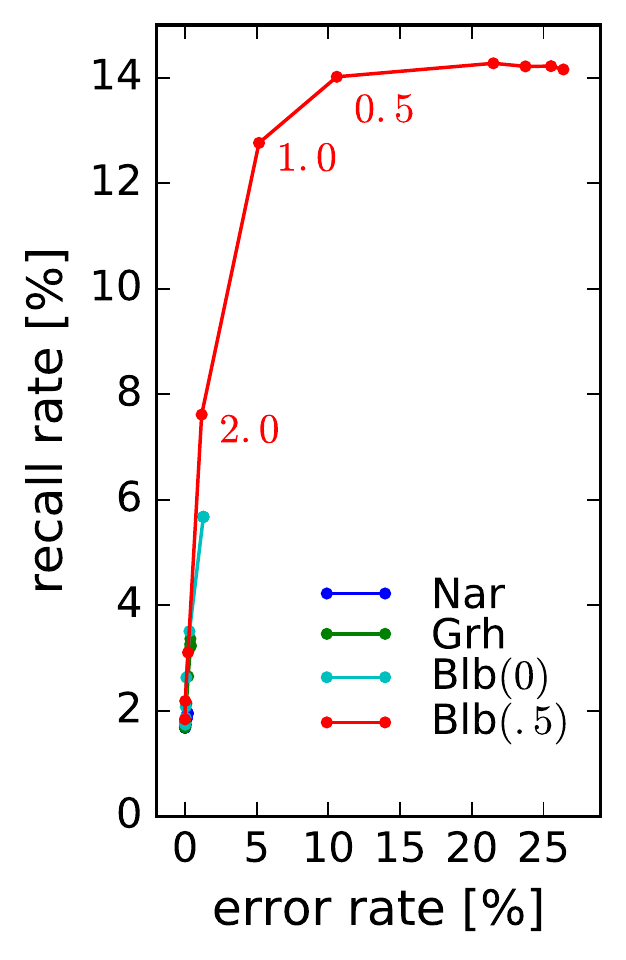}
		\caption{Different $\Theta$ values used on $\alpha_v=0.28, \alpha_e=0.09$ setting.}
		\label{fig:robustness:thetas}
    \end{subfigure}
    \caption{Robustness comparison of different algorithms (Epinions dataset); recall and error rates for different perturbation settings.}
    \label{fig:robustness}
\end{figure}

We provide results in the Epinions network in Fig. \ref{fig:robustness:perts}. A greedy setting of Bumblebee, Blb$(.5, .5)$ had the highest recall rates in all cases, with an acceptable level of error except in two settings. A more conservative setting, Blb$(.01,0)$ provided recall rates close to Nar (and higher to Grh), but with negligible error.

In experiments with small overlap, error rates of Blb$(.5,.5)$ were higher in case of all three networks and while staying low for Nar (as of no propagation); for Blb it reached $12\%$ in case of Epinions. These cases were with perturbation settings 
$\alpha_v=0.25, \alpha_e=0.75$ and $\alpha_v=0.5, \alpha_e=0.25$. However, these are the perturbation parameters, not actual values: deleting nodes delete some edges and removing edges sometime totally disconnect nodes. Therefore we displayed the actual overlaps on the x-axis in Fig. \ref{fig:robustness:perts}, which are measured with the Jaccard similarity \cite{nar09}. In these special cases the background knowledge and anonymized networks shared typically $\leq 10\%$ of overlapping edges in all three networks.

In case of $\alpha_v=0.28, \alpha_e=0.09$ only Blb$(.5,.5)$ could achieve large-scale propagation, but with high error. This can be decreased by setting parameters $\delta$ or $\Theta$. With the latter, we could achieve recall = $12.7\%$ with an error only at $5.1\%$ ($\Theta=1.0$), or recall = $7.6\%$, error = $1.1\%$ ($\Theta=2.0$); see more details in Fig. \ref{fig:robustness:thetas}.

We furthermore note that Fig. \ref{fig:robustness:thetas} also provides evidence that parameter $\Theta$ allows significantly more fine-grained control in Bumblebee than in the Nar algorithm. Fig. \ref{fig:robustness:thetas} is also an example how one can search the parameter space for the desired trade-off between recall and error by fixing one of the parameters.

\subsection{Comparison with Other Attack Algorithms}
\label{sec:comparison}
We can now compare the performance of the Bumblebee algorithm with other modern attacks. As \cite{secgraph16} provides a recent selection and comparison of efficient structural social network de-anonymization attacks, we compare Bumblebee to these attacks under the same settings. The source code of the attacks are released with the SecGraph tool \cite{secgraphweb}. Therefore, we obtained the same datasets, the Enron \cite{snap} (36k nodes, 183k edges) and Facebook \cite{koblenz} networks (63k nodes, 817k edges). With the sampling based perturbation method (implemented in \texttt{SALab}) we recreated the datasets $G_{src},G_{tar}$ as of \cite{secgraph16}: we only needed to randomly sample edges with probability $s$, while nodes were kept intact as much as possible. We note that this perturbation technique does notably smaller damage to the data compared to how we created synthetic datasets; for example, the lowest setting for Enron with $s=0.6$ corresponds to $\alpha_v=0.77, \alpha_e=0.42$, or Enron $s=0.95$ corresponds to $\alpha_v=0.97, \alpha_e=0.90$ (cf. x-axis in Fig. \ref{fig:robustness:perts}).

According to the results of Table 6 in \cite{secgraph16}, we selected top performing attacks for comparison. For brevity, we simply provide the list of these attacks without details: \emph{percolation graph matching} (later referred as YG) \cite{yg13}, \emph{distance vector matching} (DV) \cite{mob12}, \emph{reconciliation attack} (KL) \cite{kl14}. We used $50$ \texttt{top} seeding (as in \cite{secgraph16}), and run the Nar and Blb$(.1,.5)$ attacks from \texttt{SALab}. We run the three other attacks from the SecGraph implementation on the generated datasets. As both frameworks output the final the mapping $\mu$, we could safely use this for our comparison.

\begin{figure*}[ht!]
	\centering
	\includegraphics[width=\textwidth]{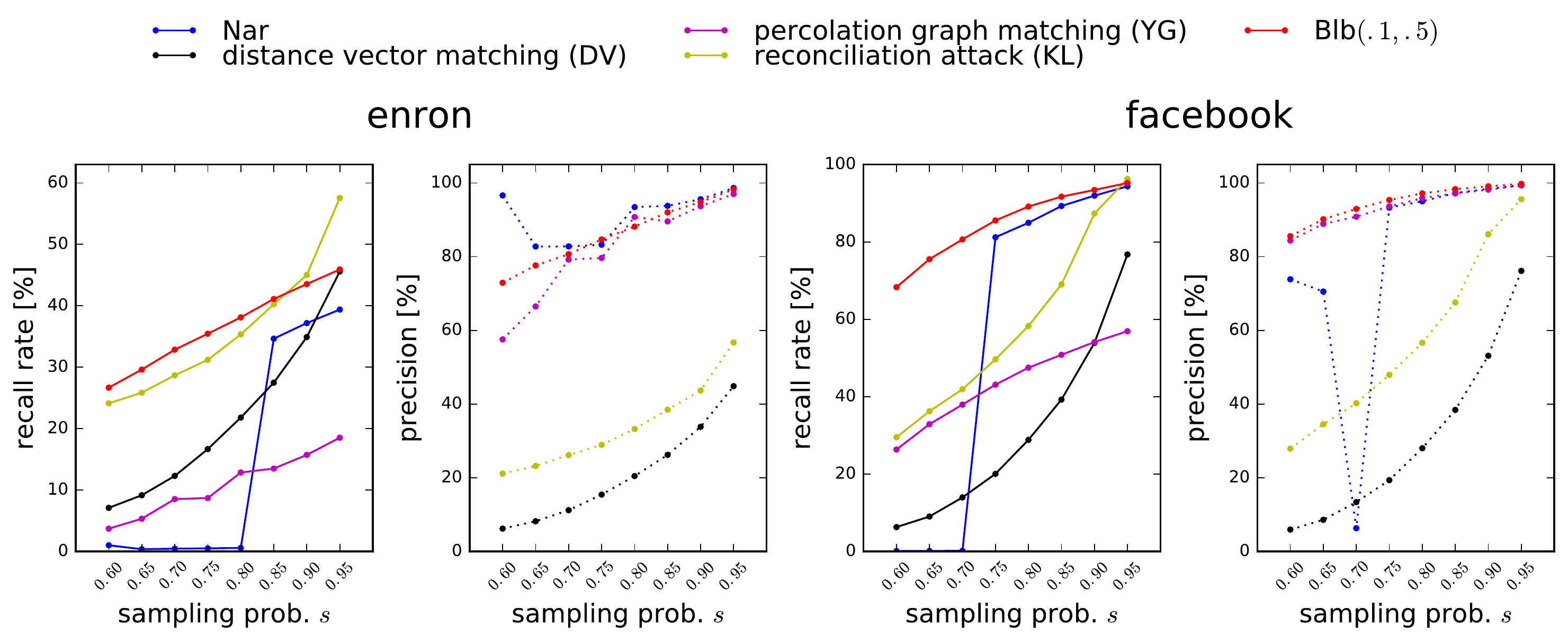}
	\caption{Comparison of efficient de-anonymization algorithms. We used the same datasets, settings and code as in \cite{secgraph16}, based on which we selected outstanding attacks in \cite{mob12, yg13, kl14} to provide a comparison with Bumblebee.}
	\label{fig:comparison}
\end{figure*}

We obtained results as in Fig. \ref{fig:comparison}. For Nar and YG we obtained almost identical results as in \cite{secgraph16}, while recall for DV were consistently smaller, and for KL significantly higher. The latter differences can be due to configurational settings
however, we tried several configurations to achieve best results for all attacks. Here, we observed that Bumblebee provided a balanced, high performance in all settings, outperforming all other attacks.

In \cite{secgraph16} only recall rates are concerned for describing the performance of de-anonymization attacks, while we found that error rates divide attacks into two strongly separated classes; we found that Nar, YG, Blb were high and DV, KL were low precision attacks. This finding puts high recalls into another perspective, as related error rates between $40$-$100\%$ are unacceptably high. Therefore, while the KL attack had higher recall rates than Blb for Enron $s=0.95$, with the $43.9\%$ error rate KL had a very poor precision, unlikely to be acceptable for an attacker.

We also evaluated robustness of attacks against anonymization schemes as a comparison of Bumblebee for results in Table 8 in \cite{secgraph16} (while keeping scheme parameters identical). Again, for brevity, we only enlist the schemes we included: \emph{random edge swithcing} (Switch) \cite{switch08}, \emph{$k$-degree anonymization} ($k$-DA) \cite{kda08}, and a method providing \emph{differential privacy} (DP) \cite{dp11}. We left out a clustering approach due to unreasonably long runtimes \cite{cluster09}, and a random walk based graph regeneration method \cite{rw13} -- against the latter we could not achieve recall rates higher than $1\%$ with any of the algorithms.

\begin{table*}[]
\centering
\caption{Robustness of attacks against different anonymization schemes from \cite{switch08, kda08, dp11}. Values are recall rates [$\%$] (denoted as Rc) and precision [$\%$] (Pr) of each algorithm under the given settings. Settings and parameters are intentionally replicating results of \cite{secgraph16} in order to enhance comparability of results.}
\label{tbl:anonymization}
\resizebox{\textwidth}{!}{ \begin{tabular}{c|c|cc|cc|cc|cc|cc|cc||cc|cc|cc|cc|cc|cc}
\multirow{3}{*}{~}    & \multirow{3}{*}{$s$} & \multicolumn{12}{c||}{Enron}   & \multicolumn{12}{c}{Facebook} \\ \cline{3-26}
    & & \multicolumn{4}{c|}{Switch($k$)} & \multicolumn{4}{c|}{$k$-DA ($k$)} & \multicolumn{4}{c||}{DP ($\epsilon$)} & \multicolumn{4}{c|}{Switch ($k$)} & \multicolumn{4}{c|}{$k$-DA ($k$)} & \multicolumn{4}{c}{DP ($\epsilon$)} \\ \cline{3-26}
    & & \multicolumn{2}{c|}{$5$} & \multicolumn{2}{c|}{$10$} & \multicolumn{2}{c|}{$5$} & \multicolumn{2}{c|}{$50$} & \multicolumn{2}{c|}{$300$} & \multicolumn{2}{c||}{$50$} & \multicolumn{2}{c|}{$5$} & \multicolumn{2}{c|}{$10$} & \multicolumn{2}{c|}{$5$} & \multicolumn{2}{c|}{$50$} & \multicolumn{2}{c|}{$300$} & \multicolumn{2}{c}{$50$} \\ \cline{3-26}
    & & Rc & Pr & Rc & Pr & Rc & Pr & Rc & Pr & Rc & Pr & Rc & Pr & Rc & Pr & Rc & Pr & Rc & Pr & Rc & Pr & Rc & Pr & Rc & Pr \\  \hline    \multirow{3}{*}{Nar} & $.85$ & $34.9$ & $94.3$ & $0.5$ & $89.9$ & $39.1$ & $98.2$ & $1.0$ & $94.6$  & $38.2$ & $98.0$ & $0.5$ & $97.3$ & $89.1$ & $96.6$ & $82.8$ & $93.6$ & $94.0$ & $98.9$ & $93.8$ & $98.8$  & $93.5$ & $98.6$ & $87.1$ & $95.8$ \\ \cline{2-26}
        & $.90$ & $36.1$ & $95.6$ & $0.7$ & $91.1$ & $40.6$ & $98.9$ & $37.7$ & $97.7$  & $39.1$ & $98.8$ & $0.5$ & $97.2$ & $90.0$ & $97.2$ & $84.4$ & $94.7$ & $94.7$ & $99.2$ & $94.6$ & $99.2$  & $94.5$ & $99.1$ & $89.7$ & $97.5$ \\ \cline{2-26}
        & $.95$ & $37.0$ & $96.5$ & $30.3$ & $94.8$ & $41.0$ & $99.2$ & $38.6$ & $98.7$  & $40.2$ & $99.4$ & $0.6$ & $98.9$ & $91.2$ & $97.9$ & $85.6$ & $95.6$ & $95.5$ & $99.7$ & $95.4$ & $99.7$  & $95.2$ & $99.6$ & $91.0$ & $98.5$ \\ \hline
    \multirow{3}{*}{YG} & $.85$ & $13.6$ & $90.0$ & $11.2$ & $86.3$ & $16.8$ & $94.7$ & $1.3$ & $24.4$  & $15.5$ & $92.0$ & $9.7$ & $76.2$ & $51.9$ & $97.8$ & $47.5$ & $96.4$ & $55.7$ & $99.0$ & $48.8$ & $92.1$  & $54.1$ & $98.3$ & $45.7$ & $89.8$ \\ \cline{2-26}
        & $.90$ & $15.0$ & $93.4$ & $12.8$ & $91.1$ & $15.2$ & $90.8$ & $1.7$ & $29.1$  & $16.9$ & $95.2$ & $10.2$ & $77.5$ & $53.4$ & $98.3$ & $49.3$ & $96.9$ & $56.6$ & $99.1$ & $52.3$ & $94.7$  & $55.2$ & $98.4$ & $48.7$ & $92.7$ \\ \cline{2-26}
        & $.95$ & $16.4$ & $94.9$ & $13.8$ & $92.1$ & $17.4$ & $95.0$ & $4.8$ & $55.1$  & $18.2$ & $96.7$ & $9.7$ & $75.8$ & $54.7$ & $98.8$ & $51.2$ & $98.2$ & $57.9$ & $99.4$ & $54.3$ & $95.9$  & $56.1$ & $98.3$ & $48.4$ & $91.5$ \\ \hline
    \multirow{3}{*}{DV} & $.85$ & $13.2$ & $13.2$ & $8.2$ & $8.2$ & $13.2$ & $13.2$ & $4.0$ & $4.0$  & $13.8$ & $13.7$ & $0.3$ & $0.3$ & $17.7$ & $17.7$ & $9.3$ & $9.3$ & $25.5$ & $25.5$ & $5.6$ & $5.6$  & $18.3$ & $18.3$ & $8.7$ & $8.7$ \\ \cline{2-26}
        & $.90$ & $15.4$ & $15.4$ & $9.4$ & $9.4$ & $15.2$ & $15.2$ & $4.8$ & $4.8$  & $14.9$ & $14.9$ & $0.3$ & $0.3$ & $22.3$ & $22.2$ & $11.7$ & $11.7$ & $31.7$ & $31.7$ & $7.1$ & $7.1$  & $23.6$ & $23.6$ & $11.1$ & $11.1$ \\ \cline{2-26}
        & $.95$ & $19.2$ & $19.2$ & $11.8$ & $11.8$ & $19.6$ & $19.6$ & $5.8$ & $5.8$  & $17.8$ & $17.7$ & $0.3$ & $0.3$ & $32.1$ & $32.1$ & $16.9$ & $16.9$ & $47.6$ & $47.6$ & $12.2$ & $12.1$  & $39.4$ & $39.4$ & $12.4$ & $12.4$ \\ \hline
    \multirow{3}{*}{KL} & $.85$ & $28.7$ & $28.7$ & $26.6$ & $26.6$ & $29.9$ & $29.9$ & $20.5$ & $20.5$  & $30.1$ & $30.1$ & $29.8$ & $29.8$ & $66.0$ & $66.0$ & $63.0$ & $63.0$ & $67.1$ & $67.1$ & $66.4$ & $66.4$  & $67.0$ & $67.0$ & $73.2$ & $73.2$ \\ \cline{2-26}
        & $.90$ & $31.2$ & $31.2$ & $28.6$ & $28.6$ & $32.9$ & $32.9$ & $23.2$ & $23.2$  & $32.2$ & $32.2$ & $34.0$ & $34.0$ & $75.3$ & $75.3$ & $71.2$ & $71.2$ & $84.9$ & $84.9$ & $84.2$ & $84.2$  & $84.0$ & $84.0$ & $81.4$ & $81.4$ \\ \cline{2-26}
        & $.95$ & $37.0$ & $37.0$ & $35.0$ & $35.0$ & $38.2$ & $38.2$ & $32.4$ & $32.4$  & $36.5$ & $36.5$ & $38.3$ & $38.3$ & $94.4$ & $94.4$ & $91.6$ & $91.6$ & $97.2$ & $97.2$ & $96.4$ & $96.4$  & $97.0$ & $97.0$ & $79.8$ & $79.8$ \\ \hline
    \multirow{3}{*}{Blb} & $.85$ & $42.5$ & $93.5$ & $38.9$ & $90.8$ & $45.2$ & $97.3$ & $42.8$ & $94.5$  & $43.8$ & $96.1$ & $27.8$ & $80.3$ & $92.1$ & $98.5$ & $88.2$ & $97.3$ & $94.8$ & $99.2$ & $94.4$ & $98.9$  & $94.5$ & $99.1$ & $89.2$ & $97.0$ \\ \cline{2-26}
        & $.90$ & $44.2$ & $95.7$ & $39.9$ & $93.2$ & $46.4$ & $98.3$ & $44.8$ & $96.7$  & $45.5$ & $97.5$ & $29.0$ & $83.3$ & $93.2$ & $99.1$ & $89.9$ & $98.4$ & $95.4$ & $99.6$ & $95.3$ & $99.5$  & $95.3$ & $99.6$ & $91.2$ & $98.2$ \\ \cline{2-26}
        & $.95$ & $45.1$ & $97.5$ & $41.2$ & $96.3$ & $47.3$ & $99.3$ & $46.1$ & $98.5$  & $46.4$ & $98.8$ & $32.0$ & $84.7$ & $93.8$ & $99.6$ & $91.0$ & $99.2$ & $96.0$ & $99.9$ & $95.9$ & $99.9$  & $95.8$ & $99.8$ & $91.9$ & $98.5$ \\ 
\end{tabular}}
\end{table*}

Here, we first sampled the background knowledge network from the original datasets with probability $s$, and obtained the anonymized network by running the anonymization algorithms also on the original datasets (no sampling). We observed robustness of evaluated attacks as in Table \ref{tbl:anonymization}. In case of anonymization, Blb proved to be generally the most robust attack, as it always achieved successful attacks with the high recall rates compared to others with high precision. It had error rates around $1$-$2\%$, exceptionally $5$-$6\%$ only for Enron, DP $\epsilon=50$. The latter was exceptional as this was the only case when another attack provided higher recall than Bumblebee; however, KL provided very poor precision with error rates between $61.7\%$-$70.2\%$.

These results show that Bumblebee is generally a balanced, robust, high performing attack with low error rates, outstanding of the state-of-the-art. Beside, while Bumblebee had runtimes typically between 5-10mins in each case, not all others were that fast, for example, KL had runtimes between 12-24h.

\subsection{Complexity and Runtime}
Runtimes of Bumblebee depend mainly on the complexity of propagation iterations and speed of convergence. The complexity of the Bumblebee propagation phase is $\mathrm{O}(d_{src} \cdot d_{tar} \cdot (|V_{src}| + 1))$, where $d_{src}, d_{tar}$ denotes maximum degree in each graph. The algorithm iterates over $\forall v \in V_{src}$. We assume having mapping $\tilde{\mu}$ when inspecting $v$, and the algorithm selects the neighbors-of-neighbors of $v$. More precisely, neighbors are $\forall v' \in V \subseteq V_{src}$ having $|V| \leq d_{src}$, and their neighbors are the neighbors of the mapped counterparts of $v$, the neighbors of $v'' = \tilde{\mu}(v')$. Here $V'' \subseteq V_{tar}$ is upper bounded as $|V''| \leq d_{tar}$. This leads us to $\mathrm{O}(|V_{src}| \cdot d_{src} \cdot d_{tar})$. We need to add the reverse checking complexity ($\mathrm{O}(d_{tar} \cdot d_{src})$) which leads us to the given bound.

However, this does not necessarily captures the total runtime, as it significantly depends on the speed on the convergence, i.e., how fast the algorithm converges to its final result in terms of the number of new re-identifications in each propagation round \cite{thesis15, grh14}. In addition, run-times need to be handled with caution and these should be only compared when Nar and Blb achieve high re-identification rates; otherwise, significant differences could occur. For instance, the number of propagation rounds could be less for the algorithm that has lower recall as it finishes before.

\begin{figure*}[ht!]
    \centering
    \begin{subfigure}[b]{0.3\textwidth}
		\centering
		\includegraphics[width=\textwidth]{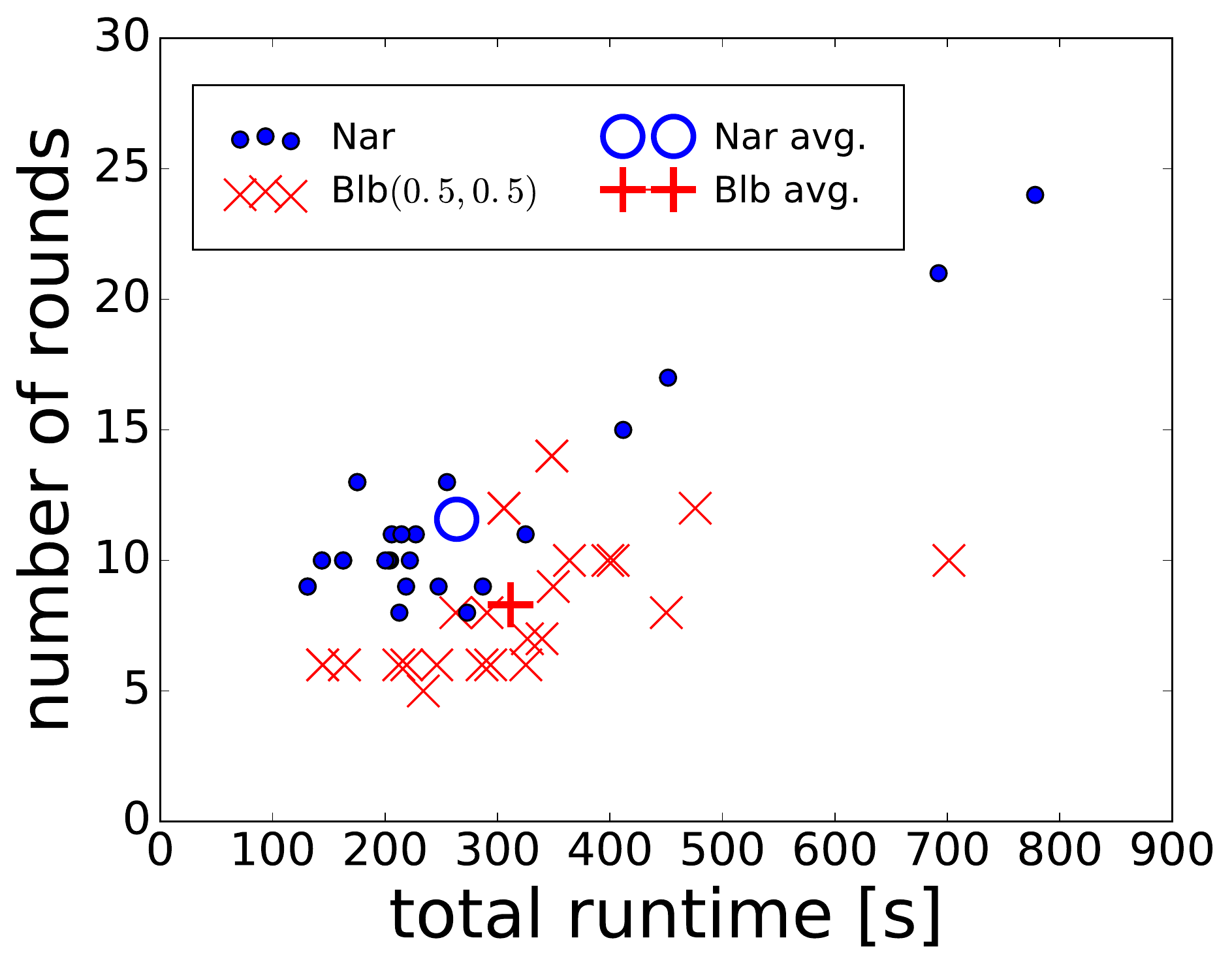}
		\caption{Number of propagation rounds and related runtimes.}
		\label{fig:runtime:numrounds}
    \end{subfigure}
    ~
    \begin{subfigure}[b]{0.3\textwidth}
    		\centering
		\includegraphics[width=\textwidth]{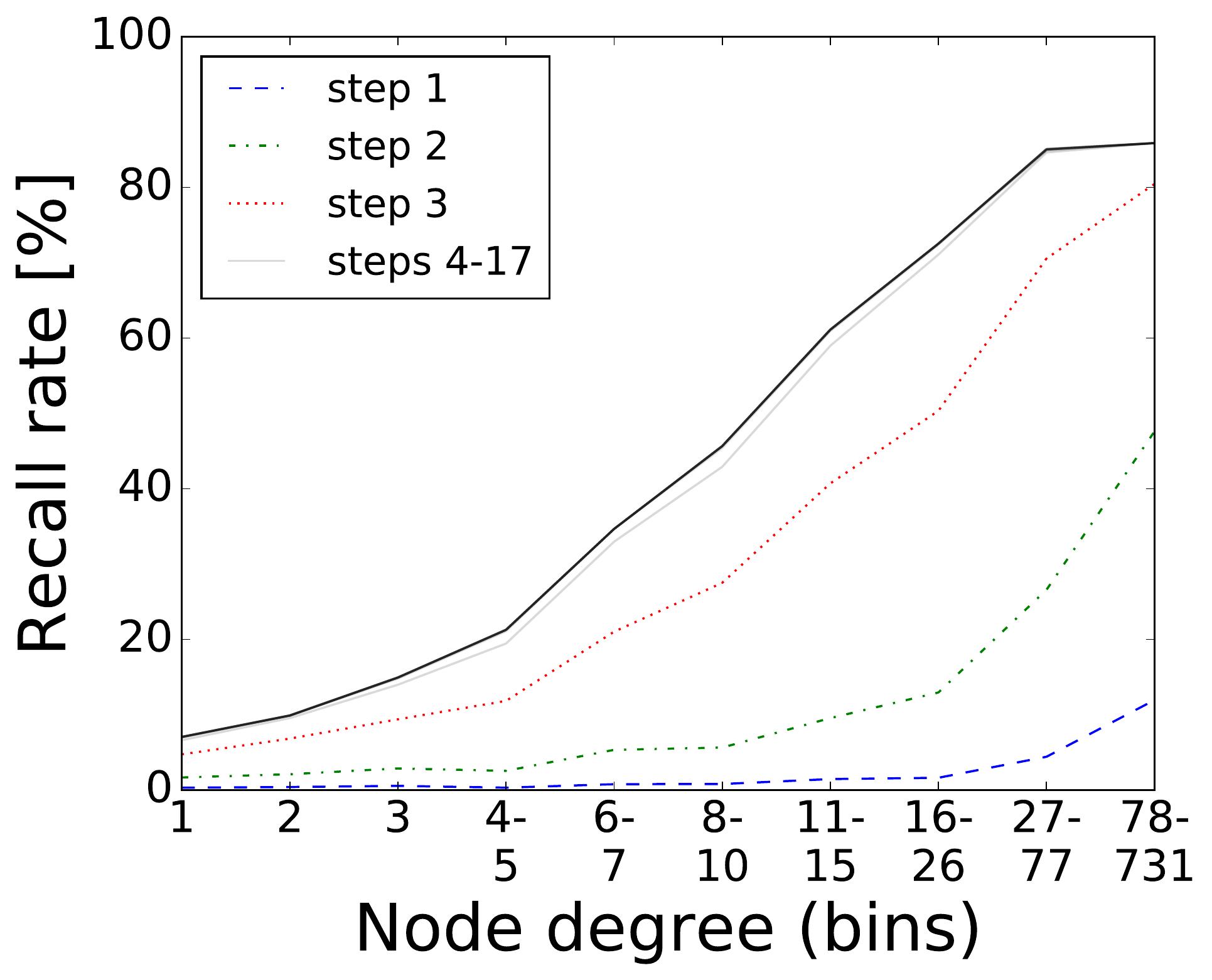}
		\caption{Re-identification convergence of Nar over degree classes.}
		\label{fig:runtime:nar}
    \end{subfigure}
    ~
    \begin{subfigure}[b]{0.3\textwidth}
    		\centering
		\includegraphics[width=\textwidth]{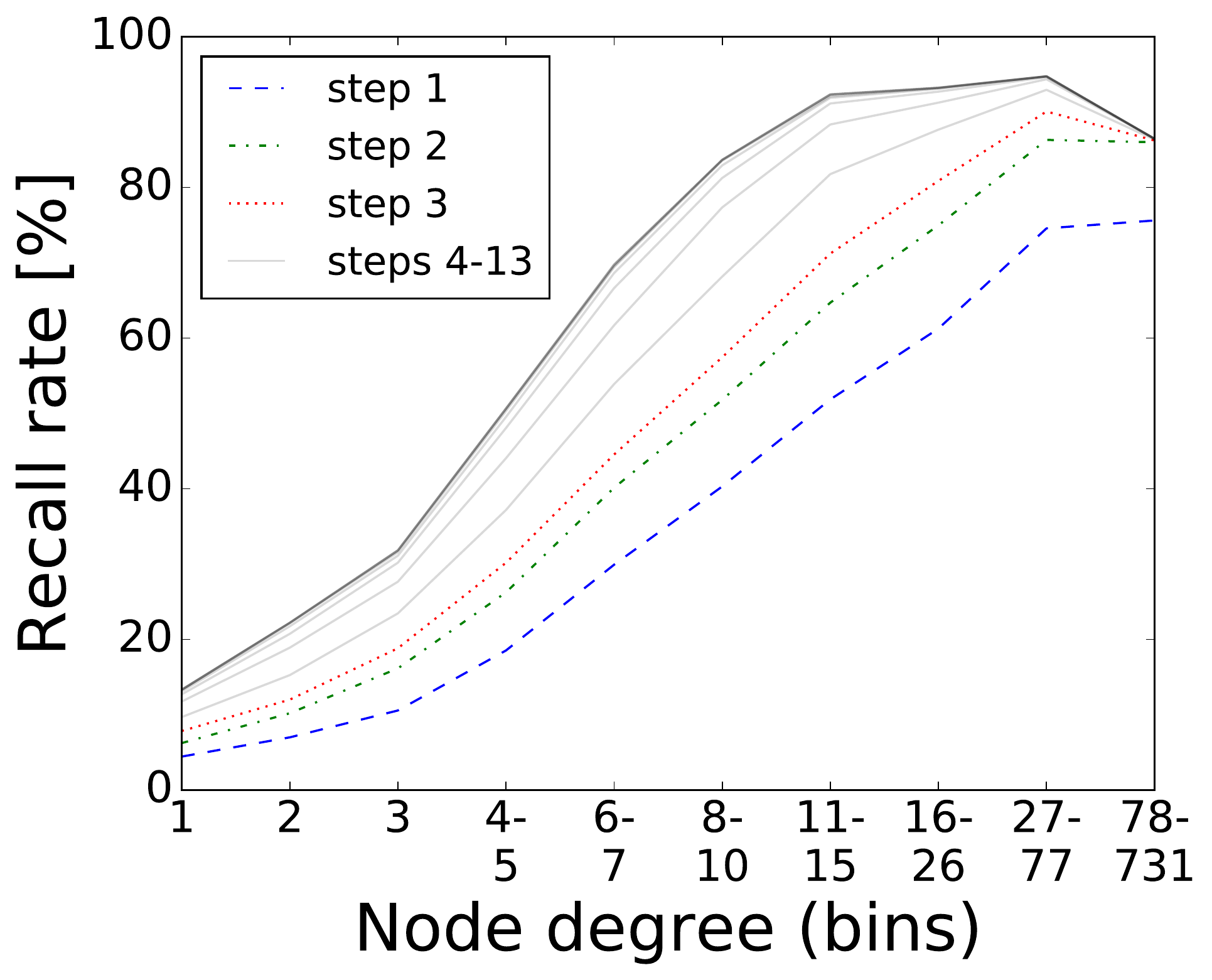}
		\caption{Re-identification convergence of Blb$(0.1,0.5)$ over degree classes.}
		\label{fig:runtime:blb}
    \end{subfigure}
    \caption{Blb converges faster in a lower number of rounds compared to Nar.}
    \label{fig:runtime}
\end{figure*}

In order to avoid such biases, we compared runtimes and the number of propagation rounds where both algorithm achieved very high recall rates both (see in Fig. \ref{fig:runtime:numrounds}). We found that Blb$(.1,.5)$ converges faster in a lower number of rounds compared to Nar, as its convergence profile is quite different (cf. Fig. \ref{fig:runtime:nar} and \ref{fig:runtime:blb}). Blb re-identifies the majority of nodes in the first propagation round and then converges for some rounds, while Nar has a slower convergence. This is also shown by the average number of propagation rounds, which was $11.5$ for Nar and $8.3$ for Blb, while each round turned to be longer for Blb. The average final runtimes were $t_{Blb} = 311$ secs and $t_{Nar} = 263$ secs, which are notable, but not dramatic differences.

\section{Conclusion}
\label{sec:conclusion}
In this paper, we considered structural social network de-anonymization attacks, where we argued that node similarity metric plays a critical part. In fact, in order to improve that critical step, we proposed a new metric designed specifically for social networks, which we incorporated our new attack called Bumblebee.

We evaluated the new attack in different real-life networks and under multiple settings. We showed that Bumblebee outperforms the baseline attack in multiple aspects, as it is robust to noise and when attacker background knowledge is weak. We have also shown improvement in the minimum required size of seed sets, achieved high recall rates. Bumblebee allows a fine-grained control over the trade-off between yield and error; which either allows fixing error rates for reaching higher recall than other attacks, or decreasing the level of error.

We have also compared Bumblebee to the other state-of-the-art attacks, and found that Bumblebee was highly efficient under all circumstances, even against different anonymization schemes. Reaching high recall with high accuracy in all cases, with balanced performance against noise and anonymization attempts make Bumblebee the most outstanding attack within the state-of-the-art.

\section*{Acknowledgements}
The authors would like to thank Amrit Kumar, Gergely {\'A}cs and Cedric Lauradoux for reviewing draft versions of this paper and for their valuable comments and discussions.
Our work is based on the idea of Benedek Simon \cite{benedekthesis15}.
This work was carried out during the tenure of an ERCIM 'Alain Bensoussan' Fellowship Programme.

\bibliographystyle{abbrv}
\bibliography{references}

\newpage

\appendix

\section{Proof for Theorem 1}

\begin{proof}
We expect that equation \eqref{eq:simexp} will hold for NarSim. Substituting equation \eqref{eq:narsim} into \eqref{eq:simexp}:

\begin{align}
\frac{|V_A|}{\sqrt{|V_A|}} > \frac{|V_A \cap V_B|}{\sqrt{|V_B|}}
\end{align}

Similarly with \eqref{eq:blbsim}, we get the following for BlbSim:

\begin{align}
|V_A| > |V_A \cap V_B| \cdot \bigg(\min \bigg(\frac{|V_A|}{|V_B|},\frac{|V_B|}{|V_A|}\bigg)\bigg)^{\delta}
\end{align}

We can rearrange these equations to the following form:

\begin{align}
  \frac{|V_A|}{|V_A \cap V_B|} > \frac{\sqrt{|V_A|}}{\sqrt{|V_B|}} \enspace\text{and}\enspace \frac{|V_A|}{|V_A \cap V_B|} > \bigg(\min \bigg(\frac{|V_A|}{|V_B|},\frac{|V_B|}{|V_A|}\bigg)\bigg)^{\delta}.
\end{align}

If NarSim scores give an upper-bound to BlbSim, then BlbSim will meet the criteria in \eqref{eq:simexp} for more cases. This means that BlbSim would make more correct similarity comparisons than NarSim.

\vfill\eject
We can write the upper-bound criteria as

\begin{align}
\label{eq:proof1}
\bigg(\min \bigg(\frac{|V_A|}{|V_B|},\frac{|V_B|}{|V_A|}\bigg)\bigg)^{\delta} \leq \frac{\sqrt{|V_A|}}{\sqrt{|V_B|}}.
\end{align}

This finally boils down to the following cases.

1. $|V_A|=|V_B|$: Here, \eqref{eq:proof1} reduces to an an equality that holds for all values of $\delta$.

2. $|V_A|>|V_B|$: We will get the following inequality

\begin{align}
\frac{(|V_B|)^{\delta}}{(|V_A|)^{\delta}} \leq \frac{\sqrt{|V_A|}}{\sqrt{|V_B|}},
\end{align}

where the left hand side is $< 1$, while the right hand side is $> 1$, thus equation \eqref{eq:proof1} holds for all values of $\delta$.

3. $|V_A|<|V_B|$: Similarly, we will get:

\begin{align}
\frac{(|V_A|)^{\delta}}{(|V_B|)^{\delta}} \leq \frac{\sqrt{|V_A|}}{\sqrt{|V_B|}}, \enspace\text{rearranged as}\enspace \bigg(\frac{|V_A|}{|V_B|}\bigg)^{\delta - 1/2} \leq 1.
\end{align}

Here, for $\delta < 1/2$, due to the negative exponent the left hand side will be $> 1$, while for $\delta \geq 1/2$ the left hand side will be $\leq 1$. 

This means, that equation \eqref{eq:proof1} holds for all three cases when $\delta \geq 1/2$, which gets us to the statement of the Theorem.
\end{proof}

\end{document}